\documentclass{aastex631}

\newcommand{\ldl}{$\lambda/\Delta\lambda$}
\newcommand{\pmra}{$\mu_{\alpha *}$}
\newcommand{\pmdec}{$\mu_{\delta}$}
\newcommand{\kms}{km\,s$^{-1}$}
\newcommand{\cms}{cm\,s$^{-2}$}
\newcommand{\masyr}{mas\,yr$^{-1}$}

\newcommand{\Msun}{M$_{\odot}$}
\newcommand{\logrphk}{$\log R^\prime_{\rm HK}$}

\newcommand{\nameSDSSshort}{SDSS~J1007+1930}
\newcommand{\name}{SDSS~J1007+1930}

\received{October 13, 2024}
\revised{\today}
\accepted{November 30, 2024}

\shorttitle{A Wide Substellar Companion to Regulus}
\shortauthors{Mamajek \& Burgasser}

\begin{document}

\title{SDSS~J100711.74+193056.2: A Candidate Common Motion Substellar Companion\\ to the Nearest B-Type Star Regulus}

\author[0000-0003-2008-1488]{Eric E.\ Mamajek}
\affiliation{Jet Propulsion Laboratory, California Institute of Technology, 4800 Oak Grove Dr., Pasadena, CA 91109, USA}
\email{mamajek@jpl.nasa.gov}

\author[0000-0002-6523-9536]{Adam J.\ Burgasser}
\affiliation{Department of Astronomy \& Astrophysics, UC San Diego, 9500 Gilman Dr., La Jolla, CA 92093, USA}
\email{aburgasser@mail.ucsd.edu}

\begin{abstract}
The L9 dwarf SDSS~J100711.74+193056.2 is situated 7${\fdg}$5 north of the nearest B-type star Regulus ($d\,=\,24.3\pm0.2$\,pc), part of a stellar quadruplet. The object is at similar distance ($d\,=\,21.9\pm1.0$\,pc) as Regulus, with a 3D separation of $3.9^{+0.6}_{-0.5}$\,pc  ($\sim$1.6 tidal radii from Regulus), and shares tangential motion within 2~\kms, hinting at a physical connection. Near-infrared spectroscopy with Keck/NIRES finds that SDSS~J100711.74+193056.2 also has a comparable radial velocity as Regulus A and B, a metallicity similar to Regulus B, and a spectral morphology consistent with the estimated 1--2~Gyr total age of Regulus's close pre-white dwarf companion. Taken together, these observations indicate that SDSS~J100711.74+193056.2 is a very widely-separated and potentially physically-bound companion to Regulus, with a binding energy and mass ratio comparable to other wide star-brown dwarf systems. It joins a growing list of brown dwarfs at the L dwarf/T dwarf transition with independent constraints on physical properties such as age and metallicity.
\end{abstract}

\keywords{L dwarfs (894) --- Multiple stars (1081) --- T dwarfs (1679) --- Wide binary stars (1801)}

\section{Introduction} \label{sec:intro}

The bright star \object{Regulus} (\object{$\alpha$ Leo}, \object{HR 3982}, \object{HD 87901}, \object{HIP 49669}) is the nearest B-type star \citep[B8IVn; $\varpi$ = $41.13\pm0.35$\, mas; $d$ = $24.31\pm0.21$\, pc; ][]{Gray03,vanLeeuwen07}, 
and the 21st brightest star in Bright Star Catalog \citep[$V$ = 1.35;][]{Hoffleit91}. It is 
part of a stellar hierarchical quadruplet with a close pre-white dwarf companion Ab, which apparently lost its envelope via mass transfer to the primary \citep{Gies20}; and two widely-separated low-mass stars (\object{HD 87884} = Regulus B \& C) \citep{Rappaport09,Gies20}. There is a wide range of ages quoted for Regulus,
with most in the range $\sim$100-250 Myr \citep[e.g.][]{David15,Stone18}. However, there is evidence that Regulus Aa has been subject to mass transfer from its proto-white dwarf companion Ab \citep{Gies20}. The progenitors of both were likely $\sim$ 1 Gyr-old dwarfs of type A \citep{Rappaport09},
an age consistent with the activity level of the K-dwarf component HD~87884 (\logrphk\, $\simeq$ $-$4.4, $\log(L_X/L_{bol})$ = $-$4.84$\pm$0.09, $\tau$ $\approx$ 0.3--2.4~Gyr; \citealt{Soderblom85,2008ApJ...687.1264M,2018A&A...616A.108B,2020ApJ...898...27S,2022A&A...661A..24F}).

While examining the proper motions of nearby stars, it was noticed that a previously known brown dwarf near Regulus had similar proper motion and parallax. \object{SDSS J100711.74+193056.2} (\object{2MASS J10071185+1930563}, hereafter \nameSDSSshort) was discovered in the Sloan Digital Sky Survey \citep{2000AJ....120.1579Y} by \citet{Chiu06}, and classified L8$\pm$1.5 based on a near-infrared (NIR) IRTF/SpeX spectrum \citep{2003PASP..115..362R}.
\citet{Schneider14} later reported an updated NIR spectral type of L9. 
Other than being a L/T transition object, \name\, has been otherwise unremarkable, although \citet{Ashraf22} recently noted a BANYAN $\Sigma$ \citep{Gagne18} membership probability of 86\%\, that it might belong to the $\sim$40-50 Myr-old \object{Argus Association}. However, this assessment was based solely on the proper motion of the source, and did not account for its 3D motion.
\nameSDSSshort\, is very faint \citep[$g$ = $25.3\pm0.6$, $z$ = $19.74\pm0.09$;][]{Ahumada20} and was not detected by Gaia \citep{2023A&A...674A...1G}.
\citet{Best20} has reported the most accurate astrometry for this object to date: $\varpi = 44.8\pm2.0$ mas ($d = 22.3\pm1.0$\,pc) and \pmra, \pmdec = $-263.2\pm1.7$, $-19.4\pm1.6$ \masyr. These parameters do not differ appreciably from those of Regulus: 
$\varpi = 41.13\pm0.35$ mas \citep{vanLeeuwen07} and \pmra, \pmdec\, = $-249.029\pm0.0455$, $6.094\pm0.0365$ \masyr\, \citep{Munn22}. 
\name\, is located nearly due north (PA$= 357{\fdg}89$) 
of Regulus at a separation of $7{\fdg}55$ (27193{\farcs}48). 
The 3D separation between the \citet{Best20} position for \name\, and the revised Hipparcos \citep{vanLeeuwen07} position for Regulus is $\Delta\,=\,3.866^{+0.607}_{-0.473}$\,pc ($798^{+125}_{-98}$\,kau; 
68.3\% confidence limit uncertainties). 
Adopting a total mass of the \object{Regulus} system of 5.46\,\Msun\ based on component masses from \citet{Rappaport09, Che11} and \citet{Gies20}, one would predict a tidal radius of $r_t \simeq 2.4$\,pc using the formula from \citet{Mamajek13}; hence, \name\, is about $\sim$1.6$r_t$ away from Regulus.

Motivated by the proximity of \name\, to Regulus and the similarity of proper motions, we report spectral observations of this brown dwarf to search for signs of youth and to measure its radial velocity (RV), both as a means to test whether \name\, might be an additional object bound to the Regulus multiple system. 
In Section~\ref{sec:obs} we report our new observations, and 
in Section~\ref{sec:analysis} we used these data to infer the physical properties of {\name} based on spectral features and fits to spectral models.
Section~\ref{sec:results} discusses the evidence of physical association between \name, and Regulus, and the potential role of the brown dwarf as a rare benchmark in the dwarf L/T transition.

\section{Observations} \label{sec:obs}

{\name} was observed twice with the Near-Infrared Echellette Spectrometer (NIRES; \citealt{2004SPIE.5492.1295W}) on the Keck II 10m telescope. Observations conducted on 27 December 2023 (UT) were obtained in clear conditions with excellent seeing (0$\farcs$4).
However, given that the seeing was smaller than the 0$\farcs$55 fixed slit width, we determined that the RV measurements could be biased by an offset center of light.
A second spectrum was obtained on 20 January 2024 (UT), again in clear conditions but with 0$\farcs$7 seeing due to gusting winds.
We obtained 4$\times$250~s exposures at an airmass of 1.01, nodding 10$\arcsec$ along the slit in an ABBA pattern, preceded by observations of the A0~V star \object{HD~96781} ($V$=10.20) for telluric absorption correction and flux calibration. Arc and flat-field lamp observations were obtained at the start of the night.
Data were reduced using a modified version of the Spextool package
\citep{2004PASP..116..362C} following standard procedures.
OH emission lines were used to determine wavelength calibration ($\sigma_\lambda$ = 0.064\,{\AA}, $\sigma_{RV} \approx 1.$\,\kms), and the A0~V star spectrum was used to correct for telluric absorption and instrumental response after \citet{2003PASP..115..389V}. 
The final data have a median signal-to-noise (S/N) ratio = 94 at the 1.27~$\micron$ peak at native resolution. 

\section{Analysis} \label{sec:analysis}

\subsection{Atmosphere Properties} \label{sec:atm}

Figure~\ref{fig} compares the full-resolution and smoothed NIRES spectrum of \name\ to that of the L9 near-infrared spectral standard \object{DENIS~J025503.3-470049} \citep{1999AJ....118.2466M,2010ApJS..190..100K}.
{\name} shows the same overall spectral energy distribution, with evidence of weak CH$_4$ absorption at 1.6~$\mu$m and 2.2~$\mu$m that argues for a T0 dwarf classification  \citep{2006ApJ...637.1067B}.
We also note that the gravity-sensitive K~I lines at 1.2432\,$\mu$m and 1.2522\,$\mu$m are clearly present, arguing against this source being a young brown dwarf \citep{1996ApJ...469..706M,2004ApJ...600.1020M,2013ApJ...772...79A}.
The near-infrared colors of this source, 
$J-K_s$ = 1.60$\pm$0.07 (MKO system; \citealt{Chiu06}),
$W1-W2$ = 0.605$\pm$0.018 (CatWISE2020; \citealt{2021ApJS..253....8M}), and
$J-W2$ = 3.22$\pm$0.05 are also in line with normal field L/T transition brown dwarfs in contrast to typically red young planetary-mass objects
\citep{2010ApJ...710.1627L,2015ApJ...810..158F,2023ApJ...943L..16S}.

We fit the smoothed and absolute flux-calibrated spectrum to a suite of model atmospheres, finding a best match to the Spectral ANalog of Dwarfs (SAND) model set \citep{2024RNAAS...8..134A}. A Markov chain Monte Carlo fitting analysis \citep{2024arXiv240708578B} yields T$_{\rm eff}$ = 1630$^{+10}_{-63}$, K, $\log{g}$ = 5.43$\pm$0.04 (\cms) and [M/H] = $-$0.13$\pm$0.02 (Figure~\ref{fig}). The temperature is warmer than typical L/T transition objects (T$_{\rm eff}$ $\approx$ 1250--1290~K; \citealt{2015ApJ...810..158F}), while
the surface gravity is consistent with a field-age brown dwarf.
The subsolar metallicity is notable, as it is similar to that measured for the K0~V star Regulus B (\object{HD~87884}), [Fe/H] = $-$0.16$\pm$0.01 \citep{2019AJ....158..101M}.
Spectral fits using other model atmospheres (e.g., ATMO, \citealt{2020AA...637A..38P}; LOWZ, \citealt{2021ApJ...915..120M}; Sonora DiamondBack, \citealt{2024arXiv240200758M}) yield similar parameters, albeit with poorer agreement with the spectral data.

\subsection{Radial Velocity} \label{sec:rv}

\begin{figure}[ht!]
\centering
\includegraphics[width=0.9\textwidth]{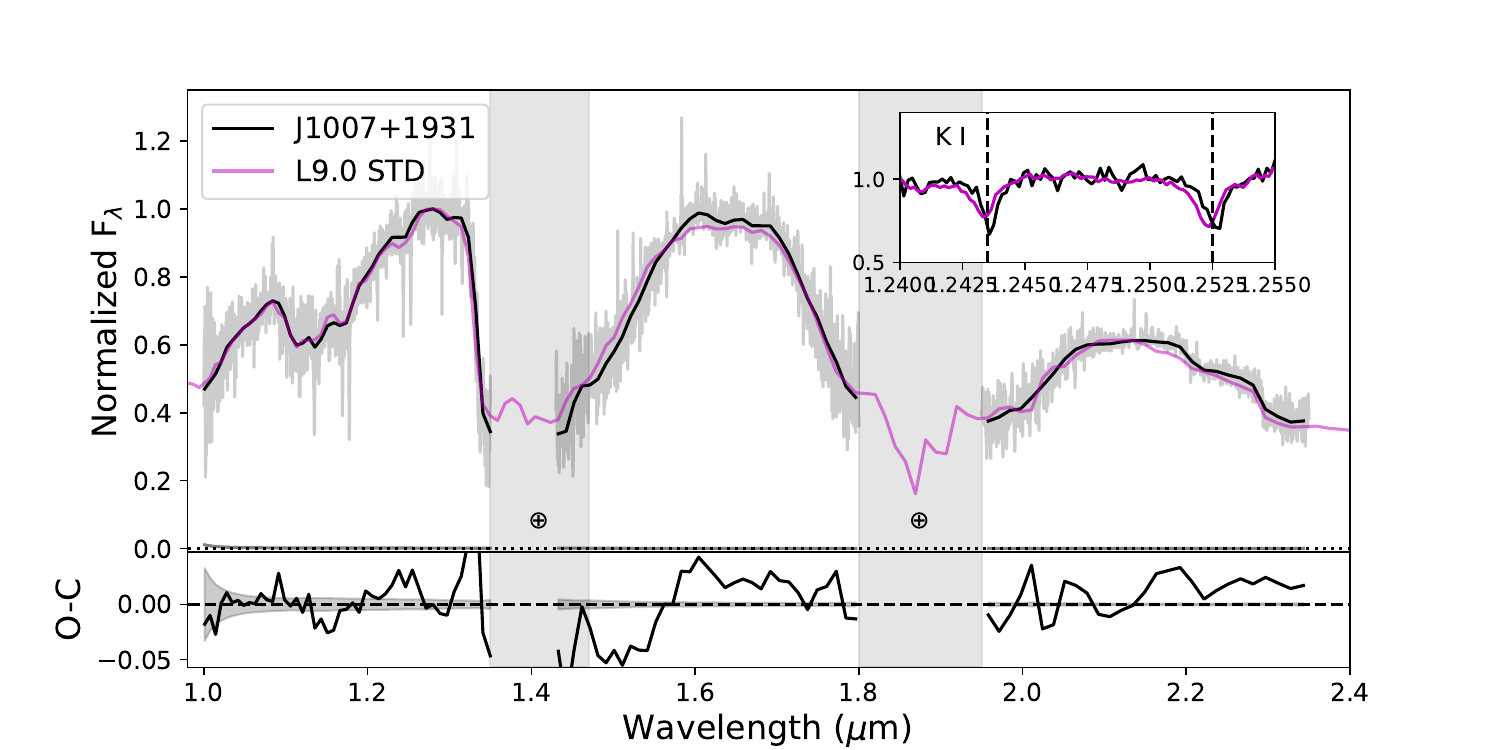} \\
\includegraphics[height=2in]{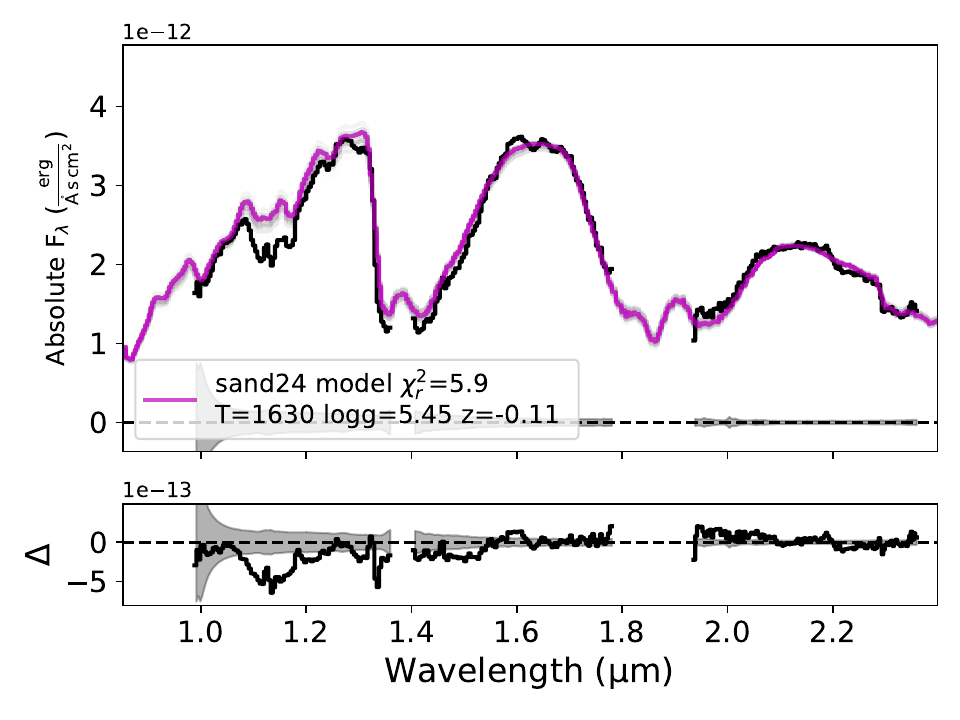}
\includegraphics[height=2in]{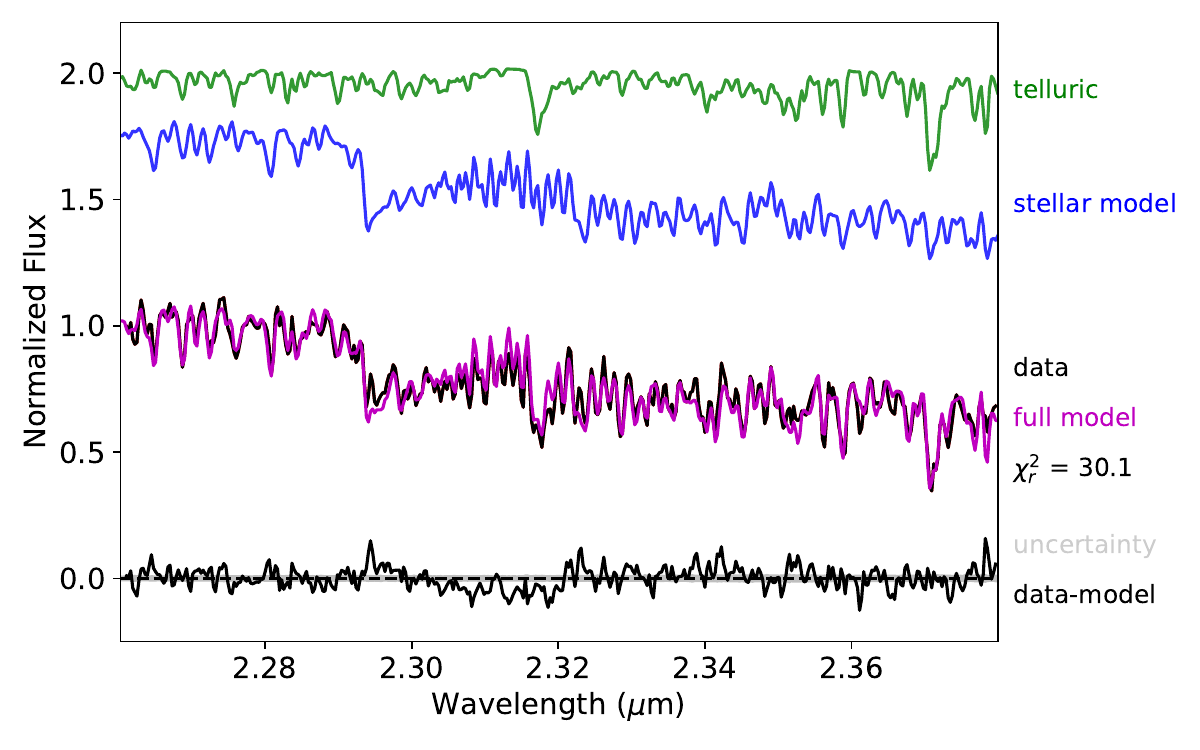}
\caption{
(Top) Keck/NIRES spectrum of \object{SDSS~J100711.74+193056.2} (grey line shows full resolution, black lines shows smoothed spectrum with {\ldl} = 100) compared to the low-resolution spectrum of the L9 near-infrared standard \object{DENIS~J025503.3-470049} (magenta line; data from \citealt{2006ApJ...637.1067B}). Regions of strong telluric absorption masked out in the data are indicated by vertical grey bands. The inset box shows a close-up of the 1.2432~$\mu$m and 1.2522~$\mu$m K~I doublet with equivalent NIRES data for \object{DENIS~J025503.3-470049} from \citet{2022RNAAS...6..151T}. Below this is the difference spectrum between \object{SDSS~J100711.74+193056.2} and \object{DENIS~J025503.3-470049} compared to the measured spectral uncertainty (horizontal grey band).
(Bottom left): Smoothed and absolute flux-calibrated NIRES spectrum of {\name} (black line) compared to the best-fit SAND model (magenta line, parameters in legend and main text), with residuals shown in the bottom panel. (Bottom right): Best-fit forward model (magenta line) of the 2.27--2.39~$\mu$m NIRES spectrum (black line), showing contributions from telluric (green line) and stellar models (blue line), with residuals (black line) shown at the bottom.
\label{fig}}
\end{figure}

The NIRES data provide sufficient resolution to infer the radial velocity of {\name}. We used a forward-modeling approach as described in \citet{2023ApJ...943L..16S}, focusing on the 2.27--2.39\,$\mu$m region of the pre-corrected spectrum of {\name} that contains both strong stellar absorption features from CO and H$_2$O, and strong telluric lines to anchor the wavelength calibration. Figure~\ref{fig} displays the best-fit model based on high-resolution Sonora Bobcat atmosphere models \citep{2021ApJ...920...85M} and the \citet{1991aass.book.....L} telluric absorption atlas. We measure a heliocentric $v_r$ = $-4\pm5$~\kms, which is formally consistent ($\lesssim$2$\sigma$) with the reported radial velocities for \object{Regulus} A 
($4.3\pm0.2$~{\kms} from \citealt{Gies08}; 0.7$\pm$0.6~{\kms} from APOGEE, 
\citealt{2020AJ....160..120J})
and \object{Regulus B} (6.6$\pm$0.2~{\kms} from \citealt{2002A&A...382..118T}; 6.59$\pm$0.13~{\kms} from Gaia DR3, \citealt{GaiaDR3}).

\section{Discussion} \label{sec:results}

Combining the measured radial velocity with astrometry from \citet{Best20}, we estimate the star's velocity vector in Galactic Cartesian frame\footnote{Galactic Cartesian frame with solar system barycenter at zero velocity, $U$ towards Galactic center, $V$ in direction of Galactic rotation ($\ell$ = 90$^{\circ}$), and $W$ towards the North Galactic Pole \citep[e.g.][]{deZeeuw1999}.} to be 
($U, V, W$) = ($-19.1\pm2.7$, $-4.8\pm1.8$, $-19.6\pm4.0$)~{\kms}. 
Combining the position and proper motion of \object{Regulus} A from \citet{Munn22}, $v_r$ = $4.3\pm0.2$ \kms\, from \citet{Gies08}, and parallax from \citet{vanLeeuwen07}, we estimate the Galactic velocity of \object{Regulus} A to be ($U, V, W$) = ($-25.12\pm0.21$, $-6.08\pm0.15$, $-13.22\pm0.19$)~{\kms}. 
The velocity components for \name and Regulus A are therefore consistent within 2$\sigma$, and the magnitude of the velocity difference is only $\left|\Delta \vec{v}\right|$ = $8.9^{+4.9}_{-4.4}$ \kms.

Adopting a brown dwarf mass of 0.06~M$_\odot$, based on the evolutionary models of \citet{2001RvMP...73..719B}, an assumed age of 1~Gyr, and T$_{\rm eff}$ = 1300~K,
the kinetic energy of {\name} in the rest frame of Regulus~A
has a 50\% probability of being $<$5$\times$10$^{43}$~erg.
In comparison, the present-day separation between these sources (the latter assumed to be the center of mass of the quadruple) corresponds to a binding energy of $\sim$10$^{40}$~erg.
Thus, it is inconclusive as to whether the system is physically bound or not. 
Indeed, the wide separation of {\name}, corresponding to a circular orbit period of about 200~Myr,
makes it susceptible to being stripped away from the Regulus system through encounters with molecular clouds or other stars. Using the formalism of \citet{1987ApJ...312..367W}, we estimate that \name\ undergoes perturbative encounters with passing stars every 1-2 orbits, with a overall diffusion timescale of $\sim$3~Gyr. 
This timescale sets a rough upper limit to the age of the wider system if it formed in its current configuration, consistent with a total system age of 1--2~Gyr. 
Alternatively, \name\ could have been ejected from a closer-in orbit through dynamical interactions among the \object{Regulus} quadruple, 
or may have formed in Regulus's proto-cluster and subsequently captured by the quadruple during the cluster's evaporation \citep{2010MNRAS.404.1835K}.
Further insight into the dynamical origin of {\name} requires a more complete orbit, which may be prohibitively long. 

The separation, binding energy, and mass ratio ($\sim$10$^{-2}$) of {\name} and Regulus A are comparable to other wide low-mass companions to nearby stars, including the quadruplet system 
\object{GJ 900} (K7+M4+M6) + CWISE J233531.55+014219.6 (T9; $\Delta$=12~kau; \citealt{2024AJ....167..253R}) and the young binary \object{L~34-26} (M3) + \object{COCONUTS~2B} (T9; $\Delta$=6.5~kau; \citealt{2021ApJ...916L..11Z}).
Despite the uncertainty of their future association, the spatial, kinematic, and age consistency between {\name} and the Regulus system implies that these sources likely formed together or within the same natal cluster.
{\name} joins a growing list\footnote{These include \object{GJ~584C} (L8, $\Delta$ = 3.5~kau; \citealt{2001AJ....121.3235K}),
\object{GJ~337CD} (L8+L8; $\Delta$ = 0.9~kau; \citealt{2001AJ....122.1989W,2005AJ....129.2849B}),
\object{BD+60~1417B} (L8$\gamma$, $\Delta$ = 1.7~kau; \citealt{2021ApJ...923...48F}),
\object{HD~46588B} (L9:, $\Delta$ = 1.4~kau; \citealt{2011ApJ...739...81L}),
\object{PM~I23492+3458B} (L9, $\Delta$ = 1.1~kau; \citealt{2014ApJ...792..119D}),
\object{NLTT~51469C} (L9; $\Delta$ = 3.8~kau; \citealt{Chiu06,2019MNRAS.487.1149G}),
and \object{$\epsilon$~Indi Ba} (T1, $\Delta$ = 1.5~kau; \citealt{2003AA...398L..29S,2010A&A...510A..99K}).} of widely-separated brown dwarf companions with independent constraints on physical properties (i.e., age from the Regulus Ab, metallicity from Regulus B), and that lie at the transition between the L dwarf and T dwarf spectral classes,
a still poorly understood phase of brown dwarf atmosphere evolution
\citep{2002ApJ...571L.151B,2006ApJ...640.1063B,2006ApJ...647.1393L,2008ApJ...685.1183L,2014ApJ...793...75R,2016ApJ...817L..19T,2021AJ....161...42B}

\begin{acknowledgments}
The data presented herein were
obtained at the W. The M. Keck Observatory is operated as
a scientific partnership between the California Institute of
Technology, the University of California, and the National
Aeronautics and Space Administration. The Observatory was
made possible by the generous financial support of the W.M.
Keck Foundation. 
The authors thank Randy Campbell, Percy Gomez, Preethi Karpoor, John Pelletier, Maxwell Piper, and Emma Softich for their assistance with the Keck observations.
The authors recognize and acknowledge the significant cultural role
and reverence that the summit of Maunakea has with the
indigenous Hawaiian community, and that the W. M. Keck
Observatory stands on Crown and Government Lands that the
State of Hawai’i is obligated to protect and preserve for future
generations of indigenous Hawaiians. 
Portions of this work were conducted at the University of California, San Diego, which was built on the unceded territory of the Kumeyaay
Nation, whose people continue to maintain their political
sovereignty and cultural traditions as vital members of the San
Diego community.
Part of this research was carried out at the Jet Propulsion Laboratory, California Institute of Technology, under a contract with the National Aeronautics and Space Administration.
\end{acknowledgments}

\vspace{5mm}
\facilities{Keck:II(NIRES)}

\software{
astropy \citep{2013AandA...558A..33A,2018AJ....156..123A,2022ApJ...935..167A},
Matplotlib \citep{2007CSE.....9...90H},
NumPy \citep{2011CSE....13b..22V},
pandas \citep{mckinney-proc-scipy-2010},
SciPy \citep{2020NatMe..17..261V},
SpeXTool \citep{2004PASP..116..362C},
SPLAT \citep{2017ASInC..14....7B}}

\clearpage

\bibliographystyle{aasjournal}

\end{document}